\documentclass{svmult}
\usepackage{multicol}

\newcommand{\xx}{{\bf x}}
\newcommand{\XX}{{\bf X}}
\newcommand{\pp}{{\bf p}}
\newcommand{\PP}{{\bf P}}
\newcommand{\SSS}{{\bf S}}
\newcommand{\mbin}{m_{\rm bin}}
\newcommand{\mtot}{m_{\rm tot}}

\newcommand{\Hint}{H_{\rm Int}}
\newcommand{\Hkep}{H_{\rm Kep}}
\newcommand{\Hjump}{H_{\rm Jump}}
\newcommand{\Hlarge}{H_{\rm Large}}
\newcommand{\Hsmall}{H_{\rm Small}}
\newcommand{\Hbint}{H_{\rm BInt}}
\newcommand{\Hbkep}{H_{\rm BKep}}
\newcommand{\Hpint}{H_{\rm PInt}}
\newcommand{\Hpkep}{H_{\rm PKep}}
\newcommand{\nbin}{N_{\rm bin}}

\newcommand{\rij}{r_{ij}}
\newcommand{\Rij}{R_{ij}}
\newcommand{\sumin}{\sum_{i=1}^N}

\begin{document}
\title*{N-Body Integrators for Planets in Binary Star Systems}
\author{John E. Chambers}
\institute{Carnegie Institution of Washington, 5241 Broad Branch Road, NW, Washington DC 20015.}
\maketitle

%
%
\section{Introduction}

The discovery of planets orbiting in binary star systems represents an exciting new field of astrophysics. The stability of planetary orbits in binary systems can only be addressed analytically in special cases, so most researchers have studied stability using long-term N-body integrations of test particles, examining binary systems with a range of masses and orbits (e.g. Wiegert and Holman 1997, Holman and Wiegert 1999, Haghighipour 2006). This has led to a good understanding of the likely regions of stability and instability in binary systems. Integrators can also been used to study the more complex problem of several finite-mass planets orbiting in a binary system, where interactions between the planets are significant. However, at the time of writing, this problem has been explored in less detail than the test-particle case, and we still lack a general theory for the stability of these systems.

N-body integrators have found a second application modelling the formation of planetary systems around binary stars (e.g. Quintana et al. 2006). Typically, these studies have paralleled those of planet formation around single stars, examining a particular stage of growth such as the formation of planetesimals, oligarchic growth, or late-stage accretion of terrestrial planets. The results of these studies are discussed extensively in the chapter in this volume by Quintana et al.

Most conventional integrator algorithms can be applied to binary star systems with little or no modification. Runge-Kutta, Bulirsch-Stoer and Everhart's RADAU integrators fall into this category for example (Press et al. 1992, Stoer and Bulirsch 1980, Everhart 1985). These algorithms contain no built-in information about the system of differential equations they are solving, so they can be applied to binary systems and single-star systems equally well.

Over the last decade and a half, symplectic integrator algorithms have become increasingly popular and are widely used to study the dynamics of planetary and satellite orbits. These algorithms have two advantages over conventional integrators. First, symplectic integrators typically have good long-term energy conversation properties. While energy is not conserved in most problems, the energy error typically makes high frequency oscillations about zero, while exhibiting no long-term trend beyond that generated by computer round-off error. Secondly, in problems involving a dominant, primary mass, such as the Sun in the Solar System, the motion of other objects about the central body can be ``built in''. A relatively small amount of computation is required to calculate the accelerations due to the central body, and a large stepsize can be used since this only needs to be small enough to resolve the perturbations between the smaller bodies. These advantages mean that symplectic integrators have become the tool of choice for many researchers at present, and they will be the focus of this chapter.

The rest of the chapter is organized as follows. Section 2 contains a review of the original mixed-variable symplectic mapping developed by Wisdom and Holman (1991). Section 3 shows how this algorithm has been modified for use specifically in binary-star systems. Section 4 shows how symplectic integrators can be improved by developing symplectic correctors, and describes a new corrector for binary-star algorithms. Section 5 discusses problems that can arise when planets come close to one or both binary stars, and what might be done to overcome these problems. Finally Section 6 contains a short summary.

%
%
\section{Mixed-Variable Symplectic Integrators}

The most widely used symplectic integrators applied to planetary systems are ``mixed-variable symplectic'' (MVS) mappings, so called because they separate a problem into two parts, each of which is solved using a different set of variables (typically  Cartesian coordinates and orbital elements). These algorithms were first introduced by Wisdom and Holman (1991) and described independently by Kinoshita et al. (1991)

To understand how these integrators work, it is easiest to start by considering Hamilton's equations for a system of $N$ bodies:
\begin{eqnarray}
\frac{dx_i}{dt}&=&\frac{\partial H}{\partial p_i} \nonumber \\
\frac{dp_i}{dt}&=&-\frac{\partial H}{\partial x_i}
\end{eqnarray}
where ${\bf x}$ and ${\bf p}$ are the coordinates and momenta of the bodies respectively, and $H$ is the Hamiltonian of the system. Using Hamilton's equations, the evolution of any quantity $q$ can be
expressed as
\begin{eqnarray}
\frac{dq}{dt}&=&\sum_{i=1}^{3N} \left(\frac{\partial q}{\partial x_i}
\frac{dx_i}{dt}+\frac{\partial q}{\partial p_i}\frac{dp_i}{dt}\right) 
\nonumber \\
             &=&\sum_{i=1}^{3N} \left(\frac{\partial q}{\partial x_i}
\frac{\partial H}{\partial p_i}-\frac{\partial q}{\partial p_i}
\frac{\partial H}{\partial x_i}\right) \nonumber \\
             &=&\{q,H\} \nonumber \\
             &=&Fq
\label{eq2}
\end{eqnarray}
where $\{,\}$ are Poisson brackets, and $F$ is an operator that depends on the Hamiltonian. The evolution of $q$ can be found by solving (\ref{eq2}), which gives
\begin{equation}
q(\tau)=e^{\tau F}q(0)
=\left(1+\tau F+\frac{\tau ^2F^2}{2}+\cdots\right)q(0)
\label{eq4}
\end{equation}

MVS integrators divide the Hamiltonian into several parts each of which can be solved efficiently in the absence of the others. Most algorithms divide $H$ into parts that can be solved analytically although this is not strictly necessary. If we separate the Hamiltonian so that $H=H_A+H_B$, with operators $A$ and $B$ corresponding to the Hamiltonians $H_A$ and $H_B$,  then
\begin{equation}
q(\tau)=e^{\tau(A+B)} q(0)
\end{equation}
where
\begin{eqnarray}
e^{\tau(A+B)}&=&1+\tau(A+B)+\frac{\tau^2(A+B)^2}{2}+\cdots \nonumber \\
          &=&1+\tau(A+B)+\frac{\tau^2(A^2+AB+BA+B^2)}{2}+\cdots
\label{eq-xy}
\end{eqnarray}
In general, the operators $A$ and $B$ will not commute so that $AB\neq BA$.

A symplectic integrator is generated by concatenating several terms of the form $\exp(a_k\tau A)$ and $\exp(b_k\tau B)$, where the $a_k$ and $b_k$ are constant coefficients. The goal is to make the resulting expression equal to (\ref{eq-xy}) up to some order in the stepsize $\tau$. This is most easily accomplished using the Baker-Campbell-Hausdorff (BCH) formula which expresses the product of two exponential operators as a single new exponential operator:
\begin{equation} 
e^Ae^B=\exp\left\{A+B+\frac{1}{2}[A,B]+\frac{1}{12}[A,A,B]
+\frac{1}{12}[B,B,A]+\cdots\right\}
\label{eq-bch}
\end{equation}
where $[A,B]=AB-BA$ and $[A,B,C]=[A,[B,C]]$ etc. (Yoshida 1990).

The most commonly used MVS algorithm is the second-order leapfrog integrator:
\begin{eqnarray}
\exp\left(\frac{\tau}{2}A\right)\exp(\tau B)
\exp\left(\frac{\tau}{2}A\right)&=&
\exp\left\{\tau(A+B)+\frac{\tau^3}{12}[B,B,A]\right. \nonumber \\
&-&\left.\frac{\tau^3}{24}[A,A,B]+O(\tau^5)\right\}
\label{eq-leapfrog}
\end{eqnarray}
This differs from the true evolution (\ref{eq-xy}) by terms proportional to $\tau^3$ and higher. Over the course of a long integration, the number of steps will be inversely proportional to $\tau$. The total error will be $O(\tau^2)$, meaning that leapfrog is a 2nd-order integrator.

Each timestep using the leapfrog algorithm consists of advancing the system corresponding to $H_A$ for a time $\tau/2$, then advancing $H_B$ for $\tau$, and finally advancing $H_A$ for $\tau/2$. The integrator consists of 3 substeps. However, the first and last of these both involve $A$, so the last substep of one timestep can be combined with the first substep of the following step. Over the course of a long integration, each timestep is effectively composed of only 2 substeps. 

Wisdom and Holman (1991) split the Hamiltonian into a part $\Hkep$ containing terms corresponding to the Keplerian motion of each planet about the central star, and a second part $\Hint$ containing direct and indirect perturbation terms due to interactions between the planets. This is accomplished using Jacobi coordinates, where the position of the innermost planet is measured with respect to the central star, and the positions of the remaining planets are measured with respect to the centre of mass of the central star and planets with lower indices. Evolution under the Keplerian part of the Hamiltonian can be calculated efficiently using Gauss's $f$ and $g$ functions (Danby 1988). If Jacobi coordinates are used, $\Hint$ is a function of the coordinates only, so this part of the problem can also be solved analytically. Jacobi coordinates are also the natural choice for hierarchical systems like the planets in the Solar System, and generally lead to smaller errors than other canonical coordinate systems, such as barycentric coordinates, for a given step size. Barycentric coordinates are an especially poor choice for the Solar System since the inner planets are less massive than the outer ones. As a result, the guiding centre for the motion of the inner planets is much closer to the Sun than the barycentre of the Solar System.

When applied to planetary systems, $\Hint \sim\epsilon \Hkep$ in the Wisdom-Holman mapping, where $\epsilon$ is the planetary to stellar mass ratio, which is typically small. The error over a long integration is therfore $O(\epsilon\tau^2)$, and the small value of $\epsilon$ ensures that MVS leapfrog performs well even though it is only a second-order integrator.

%
%
\section{Binary-Star Algorithms}

The mixed-variable symplectic (MVS) integrators described in the previous section can be applied to any hierarchical system of bodies. This means they can be used to calculate the orbital evolution of planets in a binary star system provided that the radial ordering of the planets doesn't change. The MVS algorithm can also be adapted to systems that contain multiple hierarchies, such as two binary systems in orbit about each other, by using a generalized version of Jacobi coordinates (Beust 2003). However, whenever planets come close to one another, the condition $\Hint \ll \Hkep$ is violated and MVS integrators will perform poorly. In situations where planets have eccentric orbits that cross those of their neighbours, or where close encounters are possible, the MVS algorithms must be modified.

Duncan et al. (1998) and Chambers (1999) have described two ways to do this. Here we will describe the latter approach since this leads directly to a new class of symplectic integrators that can be applied to orbits in binary systems. For reasons that will become clear, the new method requires a new set of coordinates. Ideally, these should include three spatial coordinates for the centre of mass of the system (as Jacobi coordinates do), and treat all the planets equivalently, that is, make no asumptions about their radial ordering (in contrast to Jacobi coordinates). 

Canonical heliocentric coordinates (also called democratic heliocentric coordinates) meet both of these requirements (Duncan et al. 1998). Here, the position of each planet is measured with respect to that of the central star, and the stellar coordinates are replaced with those of the centre of mass of the system:
\begin{eqnarray}
\XX_0&=&\frac{m_0\xx_0+\sum_{j=1}^Nm_j\xx_j}{\mtot} \nonumber \\
\XX_i&=&\xx_i-\xx_0
\end{eqnarray}
where subscript 0 refers to the star, $m_i$ is the mass of planet $i$, and $\mtot$ is the total mass of the system.
 
The canonically conjugate momenta (which correspond to barycentric velocities) are:
\begin{eqnarray}
\PP_0&=&\pp_0+\sum_{j=1}^N\pp_j \nonumber \\
\PP_i&=&\pp_i-\frac{m_i}{\mtot}(\pp_0+\sum_{j=1}^N\pp_j)
\end{eqnarray}

Using these coordinates, the Hamiltonian for a system of $N$ planets orbiting a single star can be split into three parts:
\begin{equation}
H=\Hkep+\Hint+\Hjump
\end{equation}
where
\begin{eqnarray}
\Hkep&=&\sumin\left(\frac{P_i^2}{2m_i}-\frac{Gm_0m_i}{R_i}\right) \nonumber \\
\Hint&=&-\sumin\sum_{j>i}\frac{Gm_im_j}{\Rij} \nonumber \\
\Hjump&=&\frac{1}{2m_0}\left(\sumin\PP_i\right)^2
\label{eq-hbinary}
\end{eqnarray}
where $R_i=|\XX_i|$ and $\Rij=|\XX_j-\XX_i|$. Note that we have dropped a term $P_0^2/2\mtot$ which simply acts to move the centre of mass at a constant velocity.

Several second-order symplectic integrators can be constructed using canonical heliocentric coordinates, for example:
\begin{equation}
\exp{\left(\frac{\tau I}{2}\right)}\exp{\left(\frac{\tau J}{2}\right)}\exp{\left(\tau K\right)}
\exp{\left(\frac{\tau J}{2}\right)}\exp{\left(\frac{\tau I}{2}\right)}
\label{eq-dhleapfrog}
\end{equation}
where $I$, $J$ and $K$ are operators associated with $\Hint$, $\Hjump$ and $\Hkep$ respectively. Other second-order algorithms are similar except that the operators are permuted, making sure that the arrangement is symmetrical in each case.

Advancing the system under $\Hint$ is straightforward since this part of the Hamiltonian is a function of the coordinates only.  As a result, the positions of all planets remain constant while the velocities change due to perturbations from the other planets. Advancing under $\Hjump$ is trivial since this is a function of the momenta only. In this case, each planet's velocity remains constant but its spatial coordinates jump by a small amount. This jump is the same for all planets, so it becomes relatively more important for objects close to the central star, a point we will return to in Section 5.  Advancing $\Hkep$ is best done using Gauss's $f$ and $g$ functions as before, noting that $\PP_i$ and $\XX_i$ are canonically conjugate.

The integrator described by (\ref{eq-dhleapfrog}) performs quite well for planetary systems in which the planets do not undergo close encounters. Typical energy errors are intermediate between those using Jacobi coordinates and barycentric coordinates for hierarchical systems, although Jacobi coordinates lose their advantage if the planets have crossing orbits. Despite consisting of 5 substeps rather than 3, the integrator described above involves slightly less computational effort than the MVS leapfrog algorithm since advancing under $\Hjump$ is trivial, while $\Hint$ only contains direct terms whereas indirect terms are also present when using Jacobi coordinates.

As with the MVS leapfrog integrator, problems arise when a pair of planets has a close encounter because $\Hint$ can become comparable in size to $\Hkep$. Chambers (1999) showed that this difficulty can be overcome using a hybrid algorithm. Here, each term in $\Hint$ is split between $\Hint$ and $\Hkep$ so that the former always remains much smaller than the latter. Under this new arrangement, the Hamiltonian is divided as follows:
\begin{eqnarray}
\Hlarge&=&\sumin\left(\frac{P_i^2}{2m_i}-\frac{Gm_0m_i}{R_i}\right)
-\sumin\sum_{j>i}\frac{Gm_im_j}{\Rij}[1-\Gamma(\Rij)] \nonumber \\
\Hsmall&=&-\sumin\sum_{j>i}\frac{Gm_im_j}{\Rij}\Gamma(\Rij) \nonumber \\
\Hjump&=&\frac{1}{2m_0}\left(\sumin\PP_i\right)^2
\label{eq-hce}
\end{eqnarray}
where $\Gamma$ is a partition function.

A second-order hybrid integrator has the same form as (\ref{eq-dhleapfrog}), that is:
\begin{equation}
\exp{\left(\frac{\tau S}{2}\right)}\exp{\left(\frac{\tau J}{2}\right)}\exp{\left(\tau L\right)}
\exp{\left(\frac{\tau J}{2}\right)}\exp{\left(\frac{\tau S}{2}\right)}
\label{eq-ce-leapfrog}
\end{equation}
where $L$ and $S$ are operators associated with $\Hlarge$ and $\Hsmall$ respectively.

The partition function is chosen so that $\Gamma(R)=1$ when $R$ is large and $\Gamma(R)\rightarrow 0$ as $R\rightarrow 0$. With this choice of $\Gamma$, it is always the case that $\Hlarge \gg \Hsmall$, and the resulting integrator remains accurate during close encounters between planets. However, $\Hlarge$ is no longer analytically soluble during a close encounter since it now includes a three-body problem. These three-body terms can be calculated using a conventional $N$-body algorithm such as Bulirsch-Stoer. Provided this is done to an accuracy level close to machine precision, the user shouldn't be able to tell in practice whether the solution was derived analytically or numerically. 

Using Bulirsch-Stoer (or any other strategy) to evolve the system through a close encounter will slow down an integration. However, only the pair of planets involved in the encounter need to be integrated in this way. All the other planets are advanced analytically under $\Hlarge$ using Gauss's functions. When there are no encounters, the algorithm becomes identical to (\ref{eq-dhleapfrog}) and there is no loss of speed.

The algorithm given by (\ref{eq-ce-leapfrog}) works well for most systems of planets orbiting a single star, and has also been used extensively to study the formation of planets from a disk of smaller bodies, since these bodies undergo many close encounters. However, the algorithm performs poorly when applied to planets orbiting in binary systems. The problem arises because $\Hjump$ now contains a momentum contribution from one of the binary stars. The stellar momentum is typically large and this leads to large changes in position for all the planets via $\Hjump$. It is no longer true that $\Hjump \ll \Hkep$, and the error per step becomes large unless an unacceptably small timestep is chosen.

As Chambers et al. (2002) have shown, the solution to this problem is to devise new coordinate systems for binary systems such that all large terms can be incorporated into a single part of the Hamiltonian. Stable planetary orbits typically fall into one of two classes: (i) those that are tightly bound to one member of a binary, or (ii) those that orbit both stars at a distance that is considerably larger than the semi-major axis of the binary orbit. Each configuration will require a different set of coordinates and we will consider the two cases separately.

%
%
\subsection{Wide Binary Case}
The Hamiltonian for a system containing $N$ planets orbiting one member of a binary star is
\begin{eqnarray}
H&=&\frac{p_A^2}{2m_A} + \frac{p_B^2}{2m_B}
  + \sumin\frac{p_i^2}{2m_i}
  - \frac{Gm_Am_B}{r_{AB}}
  - Gm_A\sumin\frac{m_i}{r_{iA}} \nonumber \\
 &-&Gm_B\sumin\frac{m_i}{r_{iB}}
  - G\sumin\sum_{j>i}\frac{m_im_j}{\rij},
\label{eq-hwide}
\end{eqnarray}
where the planets orbit star $A$ while star $B$ is a distant companion.

Making use of the hierarchical arrangement of the binary system, we define a new set of coordinates, called wide-binary coordinates, as follows:
\begin{eqnarray}
\XX_A&=&\left(\frac{m_A\xx_A+m_B\xx_B+\sum_j m_j\xx_j}{\mtot}\right),
\nonumber \\
\XX_i&=&\xx_i - \xx_A, \nonumber \\
\XX_B&=&\xx_B - \left(\frac{m_A\xx_A + \sum_j m_j\xx_j}{m_A+\sum_j m_j}\right),
\label{eq09}
\end{eqnarray}
where $m_{tot}=m_A+m_B+\sum_jm_j$ is the total mass of the system, and each of the summations run from 1 to $N$. Using these coordinates, the position of each planet is measured with respect to star $A$, while the position of star $B$ is measured with respect to the center of mass of all the other objects.

The conjugate momenta $\PP$ are:
\begin{eqnarray}
\PP_A&=&\pp_A+\pp_B+\sumin\pp_j, \nonumber \\
\PP_i&=&\pp_i - m_i\left(\frac{\pp_A+\sum_j\pp_j}
{m_A+\sum_j m_j}\right), \nonumber \\
\PP_B&=&\pp_B - m_B\left(\frac{\pp_A+\pp_B+\sum_j\pp_j}
{\mtot}\right),
\label{eq10}
\end{eqnarray}
where the summations run from 1 to $N$.

In terms of the new coordinates, the Hamiltonian can be written as
\begin{equation}
H=\Hkep + \Hint + \Hjump,
\end{equation}
where
\begin{eqnarray}
\Hkep&=&\left(\frac{P_B^2}{2\mu_{\rm bin}}
 -\frac{G\mtot \mu_{\rm bin}}{R_B}\right)
 +\sumin\left(\frac{P_i^2}{2m_i}
 -\frac{Gm_Am_i}{R_i}\right), \nonumber \\
\Hint&=&-\sumin\sum_{j>i}\frac{Gm_im_j}{R_{ij}}
 +Gm_Bm_A\left(\frac{1}{R_B}-\frac{1}{|\XX_B+\SSS|}\right)
 \nonumber \\
&+&Gm_B\sumin m_i\left(\frac{1}{R_B}-\frac{1}
{|\XX_B-\XX_i+\SSS|}\right), \nonumber \\
\Hjump&=&\frac{1}{2m_A}\left(\sumin\PP_i\right)^2
\end{eqnarray}
where $\mu_{\rm bin}=(m_A+\sum m_i)m_B/\mtot$ is the reduced mass 
of the binary system (including the mass of the planets), and 
\begin{equation}
\SSS=\frac{\sumin m_i \XX_i}{m_A+\sumin m}
\end{equation}

The terms in $\Hkep$ consist of those due to the Keplerian motion of the binary (adding the masses of the planets to star $A$) and those due to the Keplerian motion of the planets about star $A$. The terms in $\Hint$ represent the interactions between planets and also the tidal perturbations on the planets due to star $B$. Finally, $\Hjump$ contains indirect perturbation terms. 

In the absence of close encounters, $\Hint \ll \Hkep$ and $\Hjump \ll \Hkep$. Each part of the Hamiltonian can be advanced efficiently using analytic solutions. For example, the $x$ component of the acceleration of planet $k$ under $\Hint$ is given by
\begin{eqnarray}
\frac{dV_{x,k}}{dt}&=&-\frac{1}{m_k}\frac{\partial\Hint}{\partial X_k}
\nonumber \\
&=&-\left(\frac{Gm_Am_B}{m_A+\sum_i m_i}\right)\frac{X_B+S_x}
{|\XX_B+\SSS|^3} \nonumber \\
&-&\left(\frac{Gm_B}{m_A+\sum_i m_i}\right)\sumin m_i
\frac{X_B-X_i+S_x}{|\XX_B-\XX_i+\SSS|^3} \nonumber \\
&+&Gm_B\frac{X_B-X_k+S_x}{|\XX_B-\XX_k+\SSS|^3}
-\sum_{i\neq k}\frac{Gm_i}{R_{ik}^3}(X_k-X_i)
\end{eqnarray}
where ${\bf V}$ is the velocity. Note that the acceleration on planet $k$ does not involve any terms proportional to $1/m_k$, so test particles can be integrated in exactly the same way as massive planets.

The $x$ component of the acceleration on star $B$ is given by
\begin{equation}
\frac{dV_{x,B}}{dt}=Gm_A\left[\frac{X_B}{R_B^3}-\frac{X_B+S_x}
{|\XX_B+\SSS|^3}\right]
+G\sumin m_i\left[\frac{X_B}{R_B^3}-\frac{(X_B-X_i+S_x)}
{|\XX_B-\XX_i+\SSS|^3}\right]
\end{equation}

Close encounters between planets can be dealt with in the same way as for systems with a single star by partitioning the planet interaction terms between $\Hkep$ and $\Hint$ as in (\ref{eq-hce}).

One step of the new wide-binary algorithm consists of 5 substeps:
\begin{itemize}
\item Advance $\Hint$ for $\tau/2$, where $\tau$ is the timestep.
\item Advance $\Hjump$ for $\tau/2$.
\item Advance $\Hkep$ for $\tau$.
\item Advance $\Hjump$ for $\tau/2$.
\item Advance $\Hint$ for $\tau/2$.
\end{itemize}
The first and last substeps can be combined into a single substep except at the beginning of the integration or whenever output is required.

The wide-binary integrator is a second-order algorithm since the 3 pieces of the Hamiltonian are applied in a symmetric order (see Yoshida 1990). Each timestep has an error $O(\epsilon\tau^3)$, where $\epsilon$ is the ratio of the planetary mass to the stellar mass so that $\epsilon\ll 1$..

Figure 1 compares the accuracy of the wide-binary algorithm with the hybrid symplectic integrator of (\ref{eq-ce-leapfrog}) when integrating the four giant planets of the Solar System in the presence of a binary companion. The giant planets orbit the Sun, while a second solar-mass star orbits the combined system moving on an orbit with semi-major axis $a=160$ AU, eccentricity $e=0.25$ and inclination of 0. The figure shows the energy error as a function of time for a 100,000-year integration using a stepsize of 50 days. The upper panel shows the performance of the hybrid integrator, while the lower panel shows the wide-binary algorithm. The hybrid algorithm performs poorly since it treats the binary companion as the equivalent of an additional planet, so that $\Hjump$ is no longer small compared to $\Hkep$. The wide-binary algorithm, which treats the binary companion as a special body, has an energy error about three orders of magnitude lower as a result.

%
%
\subsection{Close Binary Case}
The Hamiltonian for a system of $N$ planets orbiting a close binary has the same form as (\ref{eq-hwide}) except that now it is understood that the planets orbit both members of the binary. The hierarchical nature of the system suggests we switch to the following close-binary coordinates:

\begin{eqnarray}
\XX_A&=&\frac{m_A\xx_A+\sum_j m_j\xx_j+m_B\xx_B}{\mtot}, \nonumber \\
\XX_i&=&\xx_i - \left(\nu_A\xx_A+\nu_B\xx_B\right), \nonumber \\
\XX_B&=&\xx_B - \xx_A,
\end{eqnarray}
where $\mtot$ is the total mass of all the bodies, the summations run from 1 to $N$, and
\begin{eqnarray}
\nu_A&=&m_A/\mbin \nonumber \\
\nu_B&=&m_B/\mbin
\end{eqnarray}
where $\mbin=m_A+m_B$ is the mass of the binary, 

Using these coordinates, the position of each planet is measured with respect to the center of mass of the two stars, while $\XX_B$ is the relative coordinates of the binary itself.

The conjugate momenta $P$ are
\begin{eqnarray}
\PP_A&=&\pp_A + \pp_B + \sum_{j=1}^N\pp_j, \nonumber \\
\PP_i&=&\pp_i - m_i\left(\frac{\pp_A+\pp_B
+\sum_j\pp_j}{\mtot}\right),
 \nonumber \\
\PP_B&=&{\bf p_B} - \nu_B\left(\pp_A+\pp_B\right),
\end{eqnarray}
where summation indices run from 1 to $N$.

The new Hamiltonian is
\begin{equation}
H=\Hkep + \Hint + \Hjump,
\end{equation}
where
\begin{eqnarray}
\Hkep&=&\left[\frac{P_B^2}{2\mu_{\rm bin}}
 - \frac{G\mbin\mu_{\rm bin}}{R_B}\right]
 + \sumin \left[\frac{P_i^2}{2m_i}
 - \frac{G\mbin m_i}{R_i}\right], \nonumber \\
\Hint&=&G\mbin\sumin m_i\left[\frac{1}{R_i}
-\frac{\nu_A}{|\XX_i+\nu_B\XX_B|}
-\frac{\nu_B}{|\XX_i-\nu_A\XX_B|}\right]
 \nonumber \\
&-&\sumin \sum_{j>i}\frac{Gm_im_j}{R_{ij}}, \nonumber \\
\Hjump&=&\frac{1}{2\mbin}
\left(\sumin\PP_i\right)^2
\end{eqnarray}
where $\mu_{\rm bin}=m_Am_B/(m_A+m_B)$ is the reduced mass of the binary.

Using the close-binary coordinates, terms in $\Hkep$ correspond to the Keplerian motion of the two binary stars about their common centre of mass, and of the planets about this centre of mass. In addition, $\Hint$ contains terms due to interactions between the planets, and perturbations on the planetary orbits caused by higher order moments of the binary potential. As before, $\Hjump$ contains indirect correction terms.

In the absence of close encounters, $\Hint$ and $\Hjump$ are small compared to $\Hkep$, and each part of the Hamiltonian can be advanced analytically. The $x$ component of the acceleration on planet $k$, when advancing $\Hint$, is given by
\begin{eqnarray}
\frac{dV_{x,k}}{dt}&=&\frac{G\mbin X_k}{R_k^3}
-\sum_{i\neq k}\frac{Gm_i}{R_{ik}^3}(X_k-X_i)\nonumber\\
&-&G\left[m_A\frac{(X_k+\nu_BX_B)}{|\XX_k+\nu_B\XX_B|^3}
+m_B\frac{(X_k-\nu_AX_B)}{|\XX_k-\nu_A\XX_B|^3}\right]
\end{eqnarray}
while the corresponding acceleration on star $B$ is given by
\begin{equation}
\frac{dV_{x,B}}{dt}=G\sum_i m_i\left[
 \frac{(X_i-\nu_AX_B)}{|\XX_i-\nu_A\XX_B|^3}
-\frac{(X_i+\nu_BX_B)}{|\XX_i+\nu_B\XX_B|^3}\right]
\end{equation}
where $\PP_B=\mu_{\rm bin}{\bf V}_B$

Close encounters between planets can be included in the same way as before by
dividing the planet interaction terms between $\Hkep$ and $\Hint$.

One could devise a second-order scheme using close-binary coordinates that is analogous to the second-order wide-binary integrator described above. This scheme would contain 5 terms arranged symmetrically. However, this scheme would have to use a small stepsize in order to accurately integrate the orbit of the binary star. It is more efficient to assign a separate small stepsize $\tau/\nbin$ to the binary star, and choose a
larger global sizestep $\tau$ to integrate the planets. (This is analogous to the individual-timestep procedure described by Saha and Tremaine 1994.) We can do this by splitting $\Hkep$ into a part $\Hbkep$ that involves terms in $\XX_B$ and $\PP_B$ and a part $\Hpkep$ that does not. In a similar manner, we split $\Hint$ into 2 new parts $\Hbint$
and $\Hpint$, where
\begin{eqnarray}
\Hbkep&=&\frac{P_B^2}{2\mu_{\rm bin}} - \frac{G\mbin\mu_{\rm bin}}{R_B},
 \nonumber \\
\Hpkep&=&\sum_{i=1}^{N}\left[\frac{P_i^2}{2m_i}
 - \frac{G\mbin m_i}{R_i}\right], \nonumber \\
\Hbint&=&G\mbin\sum_{i=1}^Nm_i\left[\frac{1}{R_i}
-\frac{\nu_A}{|\XX_i+\nu_B\XX_B|}
-\frac{\nu_B}{|\XX_i-\nu_A\XX_B|}\right], \nonumber \\
\Hpint&=&-\sumin \sum_{j>i}\frac{Gm_im_j}{R_{ij}}.
\end{eqnarray}

An efficient second-order close-binary scheme has the following form:
\begin{itemize}
\item Advance $\Hpint$ for $\tau/2$, where $\tau$ is the timestep.
\item Repeat the following $\nbin$ times:
\begin{itemize}
\item Advance $\Hbint$ for $\tau/(2\nbin)$.
\item Advance $\Hbkep$ for $\tau/(2\nbin)$.
\end{itemize}
\item Advance $\Hjump$ for $\tau/2$.
\item Advance $\Hpkep$ for $\tau$.
\item Advance $\Hjump$ for $\tau/2$.
\item Repeat the following $\nbin$ times:
\begin{itemize}
\item Advance $\Hbkep$ for $\tau/(2\nbin)$.
\item Advance $\Hbint$ for $\tau/(2\nbin)$.
\end{itemize}
\item Advance $\Hpint$ for $\tau/2$.
\end{itemize}

Chambers et al. (2002) suggested making $\nbin$ smaller than the global timestep by a factor equal to the ratio of the binary orbital period to the period of the innermost planet. One could also use individual timesteps for the binary companion and the planets in the wide-binary algorithm described earlier. However, the amount of computer time saved would be modest since most of the effort is required to calculate the direct perturbations between the planets and this would not change using individual timesteps.

%
%
\section{Symplectic Correctors}
Wisdom et al. (1996) showed that the performance of the MVS mapping can be improved at little extra cost by applying ``symplectic correctors''. If $\exp(M)$ represents a single step of an integrator, given by (\ref{eq-leapfrog}) for example, the addition of a corrector modifies the step to become
\begin{equation}
e^Ce^Me^{-C}
\end{equation}
where $C$ is chosen in order to remove the leading order error terms.  Over the course of multiple timesteps, the corrector and inverse corrector terms cancel out, so these only need to be applied at the beginning and end of an integration and when output is required. Thus, the addition of a corrector involves little extra computational expense over a long integration.

Following Yoshida (1990), one uncorrected step of the MVS leapfrog integrator can be expressed as
\begin{eqnarray}
e^M&=&\exp\left(\frac{\tau I}{2}\right)\exp(\tau K)
\exp\left(\frac{\tau I}{2}\right) \nonumber \\
&=&\exp\left\{\tau(K+I)+\frac{\tau^3}{12}[K,K,I]
-\frac{\tau^5}{720}[K,K,K,K,I]\right. \nonumber \\
&+&\left.O(\epsilon\tau^7)+O(\epsilon^2\tau^3)\right\}
\end{eqnarray}
where $K$ and $I$ are operators associated with $\Hkep$ and $\Hint$ respectively, $\epsilon\sim I/K$ is the planet to star mass ratio, 
and we have explicitly listed only communitator terms where $I$ occurs once.

We wish to devise a corrector that eliminates terms in $[K,K,I]$ and $[K,K,K,K,I]$. In this way we can reduce the error per step from the usual $O(\epsilon\tau^3)$ to $O(\epsilon^2\tau^3)$.

Using the identity
\begin{equation}
e^Ce^Me^{-C}=\exp\left\{M+[C,M]+\frac{1}{2}[C,C,M]+\frac{1}{6}[C,C,C,M]+
\cdots\right\}
\end{equation}
we find that a corrector of the form
\begin{equation}
e^C=\exp\left\{a\tau^2[K,I]+b\tau^4[K,K,K,I]+O(\epsilon\tau^6)+O(\epsilon^2\tau^3)\right\}
\label{eq-corrector}
\end{equation}
where $a$ and $b$ are constants, results in a corrected step
\begin{eqnarray}
e^Ce^Me^{-C}&=&\exp\left\{(K+I)\tau+\left(\frac{1}{12}-a\right)\tau^3[K,K,I]
\right. \nonumber \\
&-&\left.\left(\frac{1}{720}+b\right)\tau^5[K,K,K,K,I]+O(\epsilon\tau^7)
+O(\epsilon^2\tau^3)\right\}
\end{eqnarray}
so that the $[K,K,I]$ and $[K,K,K,K,I]$ terms can be eliminated if $a=1/12$ and $b=-1/720$.

To be useful in practice, we need to be able to express the corrector as the product of terms involving $\exp(K)$ and $\exp(I)$ separately. Following Wisdom et al. (1996), correctors of the form (\ref{eq-corrector}) can be developed by noting that
\begin{eqnarray}
Y(i,k)&=&e^{kK}e^{iI}e^{-2kK}e^{-iI}e^{kK} \nonumber \\
&=&\exp\left\{2ik[K,I]
+\frac{ik^3}{3}[K,K,K,I]+\cdots\right\}
\end{eqnarray}
where we retain only commutators in which $I$ appears once.

Combining two such expressions, we obtain a suitable series of operators for the corrector (\ref{eq-corrector}):
\begin{eqnarray}
Y(i_1,k_1)\cdot Y(i_2,k_2)&=&\exp\left\{2(i_1k_1+i_2k_2)[K,I]\right. \nonumber \\
&+&\left.\frac{1}{3}(i_1k_1^3+i_2k_2^3)[K,K,K,I]+\cdots\right\}
\end{eqnarray}
where to match (\ref{eq-corrector}), we require that
\begin{eqnarray}
i_1k_1+i_2k_2&=&\frac{1}{24} \nonumber \\
i_1k_1^3+i_2k_2^3&=&-\frac{1}{240}
\label{eq-ik-constraint}
\end{eqnarray}

There are many possible solutions to (\ref{eq-ik-constraint}), for example
\begin{eqnarray}
i_1&=&-\frac{\surd 10}{72} \nonumber \\
k_1&=&\frac{3\surd 10}{10} \nonumber \\
i_2&=&\frac{\surd 10}{24} \nonumber \\
k_2&=&\frac{\surd 10}{5}
\end{eqnarray}
which generates the symplectic corrector included in the {\it Mercury\/} integrator package (Chambers 1999). Wisdom et al. (1996) provide equivalent and higher order correctors for the alternative second-order mapping
\begin{equation}
e^M=\exp\left(\frac{\tau K}{2}\right)\exp(\tau I)
\exp\left(\frac{\tau K}{2}\right)
\end{equation}

Symplectic correctors can also be devised for the binary-star integrators described in Section 3. The problem is different in that the Hamiltonian consists of three parts rather than two, but this only complicates things slightly provided we want a corrector that eliminates only terms $O(\epsilon)$ . We start with an expression for one step of the wide or close-binary algorithms:
\begin{eqnarray}
e^M&=&\exp\left(\frac{\tau I}{2}\right)\exp\left(\frac{\tau J}{2}\right)
\exp(\tau K)\exp\left(\frac{\tau J}{2}\right)
\exp\left(\frac{\tau I}{2}\right) \nonumber \\
&=&\exp\left\{\tau(I+J+K)+\frac{\tau^3}{12}[K,K,I]
+\frac{\tau^3}{12}[K,K,J]\right. \nonumber \\
&-&\left.\frac{\tau^5}{720}[K,K,K,K,I]-\frac{\tau^5}{720}[K,K,K,K,J]
+O(\epsilon\tau^7)+O(\epsilon^2\tau^3)\right\}
\end{eqnarray}
where $I$, $J$ and $K$ are operators associated with $\Hint$, $\Hjump$ and $\Hkep$ for the wide or close-binary integrator, and we list only commutators that contain $I$ or $J$ once.

We wish to eliminate the leading-order error terms involving $[K,K,I]$ and $[K,K,J]$, as well as $[K,K,K,K,I]$ and $[K,K,K,K,J]$. This suggests we look for a corrector of the form
\begin{eqnarray}
e^C&=&\exp\left\{g\tau^2[K,I]+h\tau^2[K,J]+p\tau^4[K,K,K,I]\right. \nonumber \\
&+&\left.q\tau^4[K,K,K,J]+O(\epsilon\tau^6)+O(\epsilon^2\tau^3)\right\}
\end{eqnarray}
where $g$, $h$, $p$ and $q$ are constants, which gives a corrected step of the form
\begin{eqnarray}
&&\exp\left\{(I+J+K)\tau+\left(\frac{1}{12}-g\right)\tau^3[K,K,I]
+\left(\frac{1}{12}-h\right)\tau^3[K,K,J]  \right. \nonumber \\
&-&\left.\left(\frac{1}{720}+p\right)\tau^5[K,K,K,K,I]
-\left(\frac{1}{720}+q\right)\tau^5[K,K,K,K,J]\right.
\nonumber \\
&+&\left.O(\epsilon\tau^7)+O(\epsilon^2\tau^3)\right\}
\end{eqnarray}
so that terms in $[K,K,I]$, $[K,K,J]$, $[K,K,K,K,I]$ and $[K,K,K,K,J]$ can all be eliminated if we choose $g=h=1/12$ and $p=q=-1/720$.

Following the same procedure used to obtain the MVS corrector, we note that
\begin{eqnarray}
e^Ce^{A/2}e^Be^{A/2}e^{-C}&=&\exp\left\{A+B+[C,A]+[C,B]
+\frac{1}{2}[C,C,A]+\frac{1}{2}[C,C,B]\right. \nonumber \\
&+&\left.\frac{1}{6}[C,C,C,A]+\frac{1}{6}[C,C,C,B]
+\cdots\right\}
\end{eqnarray}
where we give only commutators that contain $A$ or $B$ once.

Hence
\begin{eqnarray}
Z(i,j,k)&=&e^{kK}e^{jJ/2}e^{iI}e^{jJ/2}e^{-2kK}e^{-jJ/2}e^{-iI}e^{-jJ/2}e^{kK} \nonumber \\
&=&\exp\left\{2ik[K,I]+2jk[K,J]+\frac{ik^3}{3}[K,K,K,I]\right. \nonumber \\
&+&\left.\frac{jk^3}{3}[K,K,K,J]+\cdots\right\}
\end{eqnarray}
retaining only commutators that contain $I$ or $J$ once.

Combining two such expressions gives
\begin{eqnarray}
&&Z(i_1,j_1,k_1)\cdot Z(i_2,j_2,k_2)=\exp\left\{2(i_1k_1+i_2k_2)[K,I]\right. 
\nonumber \\
&&\left.+2(j_1k_1+j_2k_2)[K,J]+\frac{1}{3}(i_1k_1^3+i_2k_2^3)[K,K,K,I]\right. 
\nonumber \\
&&\left.+\frac{1}{3}(j_1k_1^3+j_2k_2^3)[K,K,K,J]+\cdots\right\}
\end{eqnarray}

To provide a suitable corrector, we need to satisfy the following criteria
\begin{eqnarray}
i_1k_1+i_2k_2&=&\frac{1}{24} \nonumber \\
j_1k_1+j_2k_2&=&\frac{1}{24} \nonumber \\
i_1k_1^3+i_2k_2^3&=&-\frac{1}{240} \nonumber \\
j_1k_1^3+j_2k_2^3&=&-\frac{1}{240}
\label{eq-ijk-constraint}
\end{eqnarray}
so clearly one possible solution is
\begin{eqnarray}
i_1&=&j_1=-\frac{\surd 10}{72} \nonumber \\
k_1&=&\frac{3\surd 10}{10} \nonumber \\
i_2&=&j_2=\frac{\surd 10}{24} \nonumber \\
k_2&=&\frac{\surd 10}{5}
\end{eqnarray}

Figure 2 shows the effect of including this corrector when rerunning the integration shown in Figure 1. The upper panel shows the energy error versus time when integrating the four giant planets of the Solar System with a binary companion using the corrector. The lower panel shows the case without a corrector---the same as in Figure~1. The corrector reduces the energy error by roughly three orders of magnitude, comparable to the Jupiter/Sun mass ratio. However, because the corrector is only applied at the beginning of the integration and prior to each output, this improvement in accuracy is achieved at little computational cost.

%
%
\section{Stellar Encounters}
The binary algorithms described in the previous sections are designed to work in particular circumstances. In the close-binary algorithm, the planets are assumed to orbit the centre of mass of a binary at a distance large compared to the binary separation. If a planet comes sufficiently close to the binary stars, this assumption will no longer be valid and the algorithm will break down. In the wide-binary algorithm, planets are assumed to orbit one member of a binary, while receiving small perturbations from the other star. If the distance between a planet and its central star ever becomes comparable to the distance to the other star, the wide-binary algorithm will also break down. The accuracy of each of the algorithms depends on the hierarchy of the system being preserved. For this reason the binary integrators are unable to follow the trajectories of planets moving on transfer orbits, where the centre of a planet's motion switches from one star to another or from one star to both stars.

The wide-binary algorithm suffers from a second limitation in that a planet cannot travel too close to its central star either. If this happens, the fixed stepsize of the integrator will be too large to properly follow the planet's periastron passage and accuracy will be lost. This is a well known limitation of symplectic integrators in general, including the MVS mapping (Rauch and Holman 1999). Accuracy can be restored by regularizing the motion, so that the stepsize effectively depends on a planet's distance from the star (e.g. Preto and Tremaine 1999), but regularization becomes highly inefficient for problems involving more than a few bodies.

Levison and Duncan (2000) have suggested an alternative solution which involves a new division of the Hamiltonian such that a planet's indirect perturbation terms are added to the Keplerian part of the Hamiltonian whenever these terms become large. This is analogous to the procedure described earlier for maintaining accuracy during a close encounter between two planets. We can see how this works by considering the Hamiltonian for the single-star case described by (\ref{eq-hbinary}). In Levison and Duncan's scheme, the terms in $\Hjump$ are divided between $\Hjump$ and $\Hkep$ in such a way that the former always remains small compared to the latter:
\begin{eqnarray}
\Hlarge&=&\sumin\left(\frac{P_i^2}{2m_i}-\frac{Gm_0m_i}{R_i}\right) 
+\frac{1}{2m_0}\left(\sum_{i=1}^N\PP_i\right)^2\Lambda(R_1,R_2\ldots,R_N)
\nonumber \\
\Hint&=&-\sumin\sum_{j>i}\frac{Gm_im_j}{\Rij} \nonumber \\
\Hsmall&=&\frac{1}{2m_0}\left(\sum_{i=1}^N\PP_i\right)^2
[1-\Lambda(R_1,R_2\ldots,R_N)]
\label{eq-hce-levison}
\end{eqnarray}

Levison and Duncan (2000) advocate using a partition function of the form
\begin{equation}
\Lambda=1-\prod_{i=1}^N[1-\lambda(R_i)]
\end{equation}
where $\lambda(R)$ is chosen so that $\Lambda\rightarrow1$ when any planet approaches the star and $\Lambda=0$ when all the planets are far from the star. As with the hybrid integrator described by (\ref{eq-hce}), $\Hlarge$ has to be integrated numerically whenever $\Lambda\neq 0$. In addition, for this choice of $\Lambda$, $\Hsmall$ must also be integrated numerically. An obvious shortcoming of this procedure is that an integration will proceed slowly whenever any planet passes close to the star since all the planets have to be integrated numerically in this case. However, if close periastron passages are relatively rare, this shortcoming is not severe.

The scheme of Levison and Duncan (2000) can be applied to the wide-binary algorithm to improve the accuracy whenever a planet passes close to the central star. A similar procedure could be developed to cope with planets that stray far from the central star in a wide binary, or come close to the stars in a close binary. In either case, the planet's orbit is likely to be unstable, so this state of affairs will be short-lived, compensating for the fact that all objects will have to be integrated numerically during this stage of the evolution. 

However, this approach to dealing with stellar encounters suffers from two other drawbacks that limit its usefulness. The fact that $\Hsmall$ must be integrated numerically in addition to $\Hlarge$ can be overcome by choosing a partition function $\lambda(V)$ that depends on the planets' velocities rather than their positions. Since a planet's velocity will typically become large whenever it approaches a star, velocity can be used instead of position to identify a close encounter with a star.

A more serious problem arises when one considers low-mass planets or test particles. When advancing the system under $\Hsmall$, the rate of change of the $x$ component of velocity of planet $k$ is given by Hamilton's equations:
\begin{eqnarray}
\frac{dV_{x,k}}{dt}&=&-\frac{1}{m_k}\frac{\partial\Hsmall}{\partial X_k} \nonumber \\
&=&\frac{1}{m_k}\frac{1}{2m_0}\left(\sum_{i=1}^N\PP_i\right)^2
\left(\frac{X_k}{R_k}\right)
\frac{\lambda^\prime(R_k)}{1-\lambda(R_k)}\prod_{i=1}^N[1-\lambda(R_i)]
\end{eqnarray}

Note that the righthand side of this expression is proportional to $1/m_k$, so the rate of change of the planet's velocity will become large if the planet's mass is small, and will become infinite for massless test particles. This means the scheme cannot be used to integrate test particles, and the accuracy will be severely degraded when integrating low-mass planets. 

Unfortunately, choosing a partition function that depends on velocity rather than position does not overcome this problem. A better way to tackle this issue is to use a set of coordinates that doesn't give rise to momentum cross terms like those in $\Hjump$ or $\Hsmall$, so that indirect terms are a function of the coordinates only. Barycentric coordinates have this property, but as noted earlier, symplectic integrators that use barycentric coordinates tend to perform poorly. Jacobi coordinates won't work since they do not have the necessary properties for treating close encounters between planets.

Chambers (2003) described a new set of coordinates with the right properties to apply Levison and Duncan's scheme for integrating close encounters with a star. These coordinates, dubbed ``Yosemite coordinates'' by the author, are given by
\begin{eqnarray}
\XX_0&=&\left(\frac{m_0\xx_0+\sum_jm_j\xx_j}{\mtot}\right) \nonumber \\
\XX_i&=&(\xx_i-\xx_0)+\frac{\beta}{m_0}\sum_jm_j(\xx_j-\xx_0)
\end{eqnarray}
for the case of planets orbiting a single star, where the subscript 0 refers to the star, and 
\begin{equation}
\beta=\frac{1-\sqrt{1+\mu}}{\mu\sqrt{1+\mu}}\sim-\frac{1}{2}
\end{equation}
where
\begin{equation}
\mu=\frac{\sum_{j=1}^Nm_j}{m_0}
\end{equation}

The canonically conjugate momenta are
\begin{eqnarray}
\PP_0&=&\pp_0+\sum_{j=1}^N\pp_j \nonumber \\
\PP_i&=&\pp_i-\frac{m_i}{\mtot}\left[\frac{\pp_0+(1+\beta+\beta\mu)\sum_j\pp_j}
{1+\beta\mu}\right]
\end{eqnarray}

Using these coordinates, the Hamiltonian for a system of planets orbiting a single star is
\begin{equation}
H=\Hkep+\Hint
\end{equation}
where
\begin{eqnarray}
\Hkep&=&\sum_{i=1}^N\left(\frac{P_i^2}{2m_i}-\frac{Gm_0m_i}{R_i}\right) \nonumber \\
\Hint&=&-\sum_{i=1}^N\sum_{j>i}\frac{Gm_im_j}{R_{ij}}
+\sumin Gm_0m_i\left(\frac{1}{R_i}-\frac{1}{|\XX_i-{\bf S}|}\right)
\end{eqnarray}
where
\begin{equation}
{\bf S}=\frac{\beta\sum_jm_j\XX_j}{m_0(1+\beta\mu)}
\end{equation}

The second set of terms in $\Hint$ represent indirect perturbations on each planet. These terms can be divided between $\Hint$ and $\Hkep$ in an analogous manner to the scheme of Levison et al. (2000) as follows:
\begin{eqnarray}
\Hlarge&=&\sum_{i=1}^N\left(\frac{P_i^2}{2m_i}-\frac{Gm_0m_i}{R_i}\right) \nonumber \\
&+&\sumin Gm_0m_i\left(\frac{1}{R_i}-\frac{1}{|\XX_i-{\bf S}|}\right)
\Lambda(R_1,R_2,\ldots,R_N) \nonumber \\
\Hsmall&=&-\sum_{i=1}^N\sum_{j>i}\frac{Gm_im_j}{R_{ij}} \nonumber \\
&+&\sumin Gm_0m_i\left(\frac{1}{R_i}-\frac{1}{|\XX_i-{\bf S}|}\right)
[1-\Lambda(R_1,R_2,\ldots,R_N)]
\label{eq-hce-yosemite}
\end{eqnarray}
where $\Lambda$ is chosen so that $\Lambda\rightarrow 1$ whenever any planet approaches the star and $\Lambda=0$ when all planets are far from the star.

The advantage of using Yosemite coordinates now becomes clear: each of the partitioned indirect terms in (\ref{eq-hce-yosemite}) is proportional to $m_i$. As a result, low-mass planets and test particles can be integrated to the same accuracy as massive planets without the problems encountered when using (\ref{eq-hce-levison}). The same scheme can be extended to binary systems by using new coordinate systems analogous to Yosemite coordinates with the binary hierarchy built in, as in the wide-binary and close-binary coordinate systems.

\section{Summary}
Conventional integration algorithms such as Runge-Kutta and Bulirsch-Stoer can be applied to planetary systems orbiting binary stars with little or no modification. However, these algorithms tend to be slow and exhibit long-term growth in energy errors. For these reasons, symplectic integration algorithms have become the tool of choice for many researchers. In this chapter, I have described how the standard mixed-variable symplectic map (MVS) of Wisdom and Holman (1991) can be adapted for use with planets in binary systems, including modifications to handle close approaches between planets with each other and with the stars themselves. In addition, I have shown that the performance of these algorithms can be improved at little extra cost by using symplectic correctors. These adaptations mean that symplectic algorithms can now be applied to a wide variety of problems involving planets in binary-star systems. However, there is still scope for future improvements. In particular, the current method for following close encounters between a planet and a star is slow and cumbersome, and there remains no easy way to handle planets whose orbital motion switches from one star to the other.

%
%

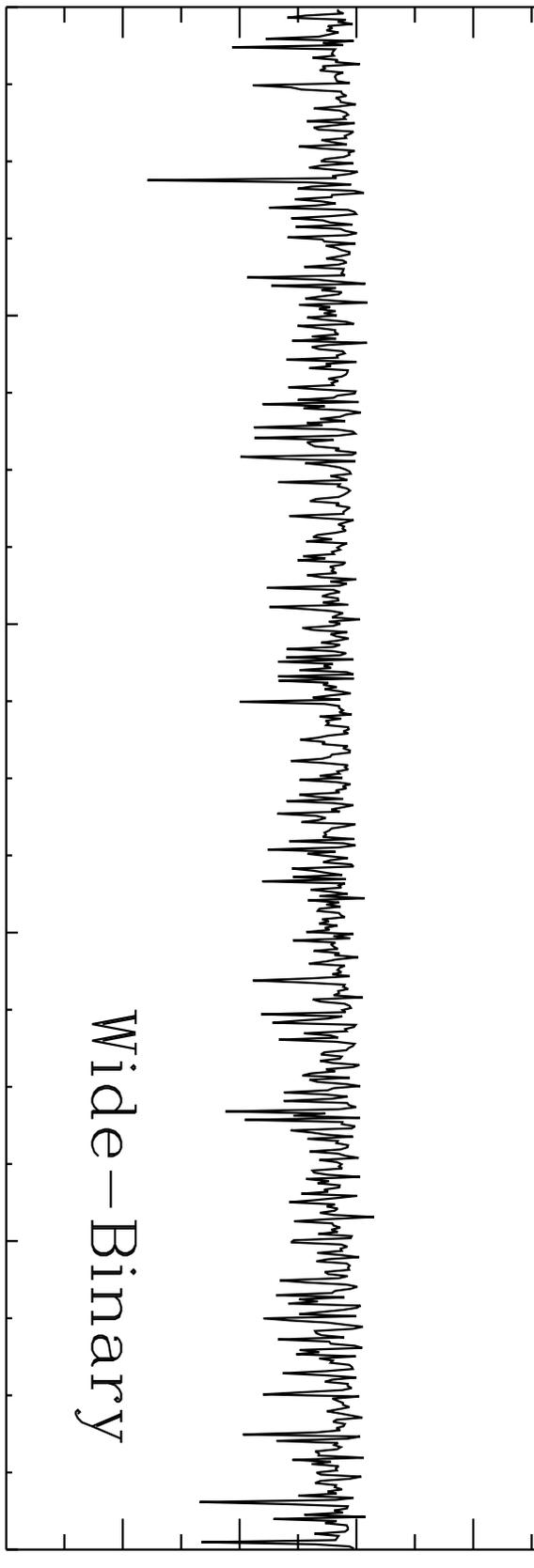
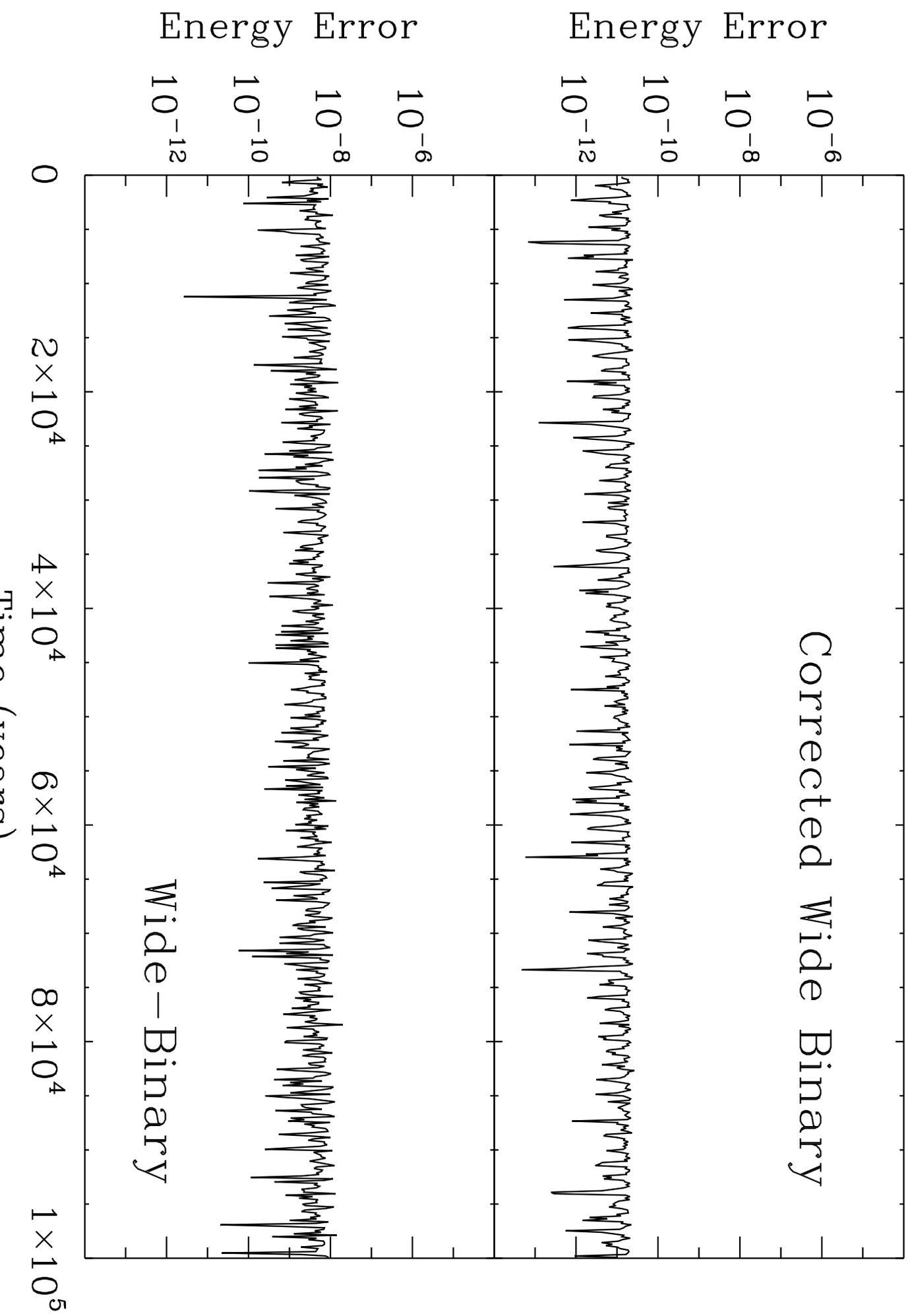
\end{document}